\documentclass[aps,prb,twocolumn,showpacs,preprintnumbers,amsmath,amssymb,superscriptaddress]{revtex4}%

\usepackage{graphicx}%
\usepackage{dcolumn}
\usepackage{amsmath}
\usepackage{color}

\begin{document}


\title{Coexistence of incommensurate magnetism and superconductivity in Fe$_{1+y}$Se$_x$Te$_{1-x}$}

\author{R.~Khasanov}
 \email{rustem.khasanov@psi.ch}
 \affiliation{Laboratory for Muon Spin Spectroscopy, Paul Scherrer
Institut, CH-5232 Villigen PSI, Switzerland}
\author{M.~Bendele}
 \affiliation{Laboratory for Muon Spin Spectroscopy, Paul Scherrer
Institut, CH-5232 Villigen PSI, Switzerland}
 \affiliation{Physik-Institut der Universit\"{a}t Z\"{u}rich,
Winterthurerstrasse 190, CH-8057 Z\"urich, Switzerland}
\author{A.~Amato}
 \affiliation{Laboratory for Muon Spin Spectroscopy, Paul Scherrer
Institut, CH-5232 Villigen PSI, Switzerland}
\author{P.~Babkevich}
 \affiliation{Department of Physics, Clarendon Laboratory, Oxford University,
 Oxford OX1 3PU, United Kingdom}
\author{A.T.~Boothroyd}
 \affiliation{Department of Physics, Clarendon Laboratory, Oxford University,
 Oxford OX1 3PU, United Kingdom}
\author{A.~Cervellino}
 \affiliation{Swiss Light Source, Paul Scherrer Institut, CH-5232 Villigen, Switzerland } %
\author{K.~Conder}
 \affiliation{Laboratory for Developments and Methods, Paul Scherrer Institute,
CH-5232 Villigen PSI, Switzerland}
\author{S.N.~Gvasaliya}
 \affiliation{Laboratory for Neutron Scattering, Paul Scherrer Institute and ETH
 Z\"urich,CH-5232 Villigen PSI, Switzerland}
\author{H.~Keller}
 \affiliation{Physik-Institut der Universit\"{a}t Z\"{u}rich,
Winterthurerstrasse 190, CH-8057 Z\"urich, Switzerland}
\author{H.-H.~Klauss}
 \affiliation{IFP, TU Dresden, D-01069 Dresden, Germany}
\author{H.~Luetkens}
 \affiliation{Laboratory for Muon Spin Spectroscopy, Paul Scherrer
Institut, CH-5232 Villigen PSI, Switzerland}
\author{V.~Pomjakushin}
 \affiliation{Laboratory for Neutron Scattering, Paul Scherrer Institute and ETH
 Z\"urich,CH-5232 Villigen PSI, Switzerland}
\author{E.~Pomjakushina}
 \affiliation{Laboratory for Developments and Methods, Paul Scherrer Institute,
CH-5232 Villigen PSI, Switzerland}
\author{B.~Roessli}
 \affiliation{Laboratory for Neutron Scattering, Paul Scherrer Institute and ETH
 Z\"urich,CH-5232 Villigen PSI, Switzerland}
%
%

\begin{abstract}
We report an investigation into the superconducting and magnetic
properties of Fe$_{1+y}$Se$_{x}$Te$_{1-x}$ single
crystals by magnetic susceptibility, muon spin rotation,
and neutron diffraction. We find three regimes of
behavior in the phase diagram for $0\leq x\leq 0.5$: (i)
commensurate magnetic order for $x\lesssim 0.1$, (ii) bulk
superconductivity for $x \sim 0.5$, and (iii) a range $0.25\lesssim
x\lesssim 0.45$ in which superconductivity coexists with static
incommensurate magnetic order.  The results are qualitatively
consistent with a two-band mean-field model in which itinerant
magnetism and extended $s$-wave superconductivity are competing
order parameters.
\end{abstract}

\maketitle


The recently discovered Fe-based high-temperature superconductors
(HTS) host an intriguing competition between magnetic, structural
and superconducting phases.  The parent phases, such as LnOFeAs
(Ln1111, Ln=La, Ce, Pr, Sm) \cite{Cruz08,Klauss08,Luetkens09,Chen08,
Zhao08,Zhao08_2, Rotundu09,Carlo09,Drew09,Sanna09} and AFe$_2$As$_2$
(A122, A = Ba, Sr, Ca),
\cite{Jesche08,Aczel08,Huang08,Zhao08_3,Goldman08} exhibit
commensurate, static, magnetic order. Upon doping or application of
pressure (chemical or mechanical), magnetism  is suppressed and
superconductivity emerges in a manner somewhat dependent on the
material.
Experiments on fluoride-doped La1111 and Pr1111 indicate
that the transition from the superconducting to the magnetic state
is of the first-order.\cite{Luetkens09,Rotundu09} Ce1111 shows a behavior which is more consistent with a quantum-critical point
separating magnetic and superconducting states.\cite{Zhao08} The
experiments on Sm1111 and A122 demonstrate co-existence of magnetism
and superconductivity.\cite{Aczel08,Drew09,Sanna09,Laplace09,Park09}

Recently, Sales {\it et al.}\cite{Sales09} reported on the synthesis
of large single crystals of Fe$_{1+y}$Se$_{x}$Te$_{1-x}$ ($0\leq x
\leq 0.5$) belonging to the 011 family of Fe-based HTS. Resistivity
measurements showed traces of superconductivity at $T\lesssim 14$~K
for all $x\neq 0$ crystals, while bulk superconductivity was
detected only for compositions close to $x=0.5$. Non-superconducting
Fe$_{1+y}$Te with $y\lesssim 0.1$ exhibits long-range commensurate
magnetic order,\cite{Fruchart75,Li09,Bao09} but only short-range
incommensurate magnetism survives in Se-doped samples.\cite{Li09,Bao09,Wen09,Lumsden09}  Up to now, however, the relation between the magnetic
and superconducting properties has not been
studied systematically for this new system. 
Here we report on a detailed study of the
evolution of superconducting and magnetic
properties of Fe$_{1+y}$Se$_{x}$Te$_{1-x}$ single crystals through a combination
of magnetic susceptibility, muon spin rotation,
and neutron diffraction. At the boundary between
magnetic and superconducting phases we observe a region of doping in
which superconductivity coexists with incommensurate magnetic order.
The phase diagram is qualitatively consistent with a two-band
itinerant models of the Fe pnictides in which magnetism and extended
$s$-wave superconductivity are competing orders.\cite{Vorontsov09,Cvetkovic09}


Muon-spin rotation ($\mu$SR) and neutron diffraction (powder and
single crystal) experiments were performed on the $\pi$M3 and
$\pi$E1 beam lines at S$\mu$S, and on the HRPT and TASP instruments
at SINQ (all at the Paul Scherrer Institute, Switzerland).
AC susceptibility measurements were performed on a
Quantum Design PPMS magnetometer with a measuring field $\mu_0H_{\rm
AC}=0.1$~mT and frequency $\nu=1000$~Hz. To reduce the effects of
demagnetization thin plate-like pieces of
Fe$_{1+y}$Se$_{x}$Te$_{1-x}$, cleaved from the main single crystals,
were oriented with the flat surface ($ab$ plane) parallel to the AC
field.

Single crystals of Fe$_{1+y}$Se$_{x}$Te$_{1-x}$ were grown by a
modified Bridgeman method similar to that reported in Ref.~\onlinecite{Sales09}. Powders of Fe, Se and Te of minimum purity
99.99\% were mixed in the appropriate ratios, pressed into a rod and
vacuum sealed in a double-walled quartz ampule. The rod was first
melted and homogenized at 1200$^{\rm o}$C for 4 hours and then
cooled in a temperature gradient 8$^{\rm o}$C/cm at a rate 4$^{\rm
o}$C/h down to 750$^{\rm o}$C followed by 50$^{\rm o}$C/h cooling.
%
Several of the crystals were ground into a powder and their phase
purity was checked by neutron powder diffraction. The amount of the
main ($P4/nmm$) fraction was found to be $\simeq 94$\%, 97\%, 98\%,
and 99\% for $x=0.5$, 0.45, 0.4, and 0.25 crystals, respectively.

\begin{figure}[htb]
\includegraphics[width=0.65\linewidth]{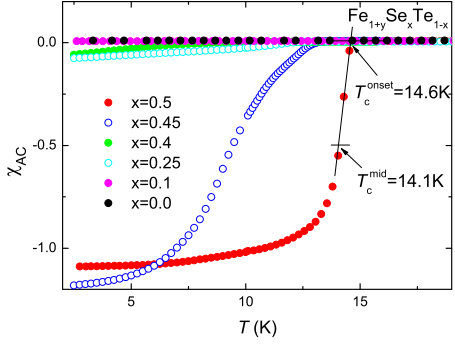}
 \vspace{-0.3cm}
\caption{(Color online) Temperature dependence of the AC
volume susceptibility $\chi_{\rm AC}$ of Fe$_{1+y}$Se$_{x}$Te$_{1-x}$.
The onset $T_{\rm c}^{\rm onset}$ and mid-point $T_{\rm c}^{\rm
mid}$ of the superconducting transition are determined from the
intersection of straight lines fit to the data above and below the
transition, and the point corresponding to $\chi_{\rm AC}=-0.5$,
respectively.}
 \label{fig:magnetization}
\end{figure}
The AC susceptibility ($\chi_{\rm AC}$) data are shown in
Fig~\ref{fig:magnetization}.
The $x=0.5$ and $x=0.45$ samples are seen to be bulk superconductors
with $\chi_{\rm AC}=-1.09$ and $-1.18$, respectively, at
$T\simeq2$~K. Values of $|\chi_{\rm AC}|$ in excess of unity are
likely explained by small non-zero demagnetization factors caused by
slight misalignment of the crystals relative to the direction of the
AC field. The $x=0.4$ and $0.25$ samples exhibit superconductivity
but have a small superconducting fraction of order 10\% at low
temperature. No traces of superconductivity were detected for the
$x=0.1$ and 0.0 samples.

$\mu$SR experiments in zero magnetic field (ZF), transverse field
(TF) and longitudinal field (LF), were performed to study the
magnetic response of the samples. In TF experiments muons stopping
in magnetically ordered parts of the sample lose their polarization
relatively fast, since the magnetic field at the muon stopping site
becomes a superposition of the external and the internal fields. The
experiments in ZF provide information on the internal magnetic field
distribution, while complementary LF measurements make it possible
to discriminate between static and fluctuating fields.

\begin{figure}[htb]
\includegraphics[width=1.0\linewidth]{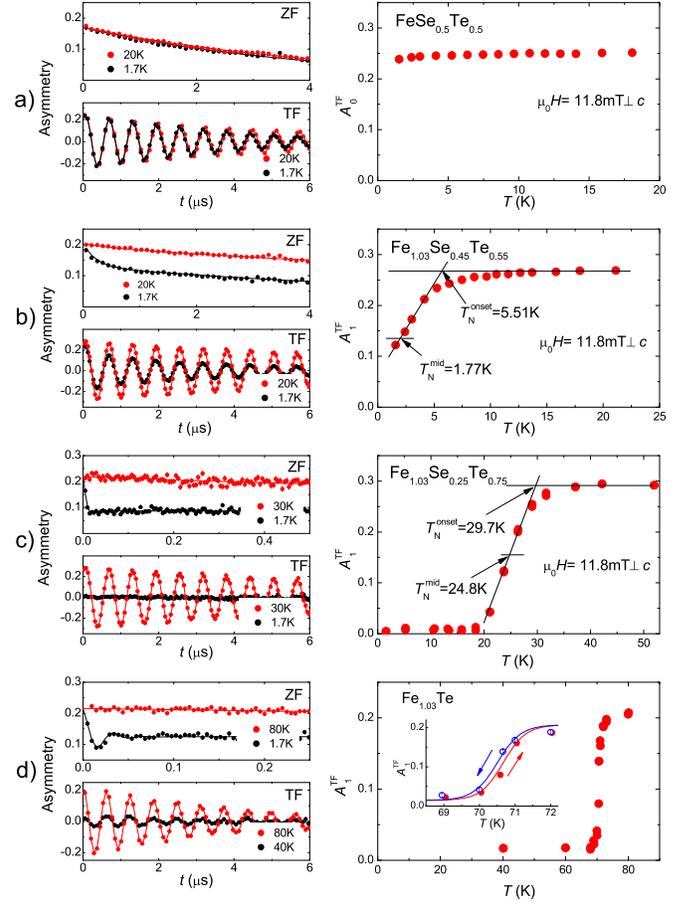}
 \vspace{-0.5cm}
\caption{(Color online) Representative ZF and TF $\mu$SR time
spectra (left panels) and temperature-dependent initial TF asymmetry of the
slow relaxing component ($A_0^{\rm TF}$ and $A_1^{\rm TF}$, right panels) for single crystals of
Fe$_{1+y}$Se$_{x}$Te$_{1-x}$. The onset ($T_{\rm N}^{\rm
onset}$) and the mid-point ($T_{\rm N}^{\rm mid}$) of the magnetic transition are determined from the
intersection of straight lines fit to the data above and below the
transition and as the point where the asymmetry decreases by a factor of 2 from its maximum value, respectively. }
 \label{fig:muSR}
\end{figure}

Figure~\ref{fig:muSR} represents $\mu$SR results for
representative compositions of Fe$_{1+y}$Se$_{x}$Te$_{1-x}$.
Figure~\ref{fig:muSR}(a) shows that for $x=0.5$ there is no
difference between the ZF time-spectra measured at $T=1.7$~K and
20~K. This suggests that the magnetic state of
FeSe$_{0.5}$Te$_{0.5}$ is the same above and below the
superconducting transition temperature. The solid lines correspond
to a fit by the function $A^{\rm ZF}(t)=A_0^{\rm ZF}
e^{-\Lambda^{\rm ZF} t}$, where $A_0^{\rm ZF}$ is the initial
asymmetry and $\Lambda^{\rm ZF}$ is the exponential relaxation
rate. Measurements in LF geometry (not shown) indicate that the
exponential character of the muon-spin relaxation is due to
randomly-oriented local magnetic fields, which are static on the
$\mu$SR time scale. Such behavior is consistent with dilute Fe
moments as observed recently for another representative of
Fe-based HTS FeSe$_{1-x}$.\cite{Khasanov08_FeSe} The TF data for
$x=0.5$ fit well to the function $A^{\rm TF}(t)=A_0^{\rm
TF}e^{-(\Lambda^{\rm TF} t+\sigma^2t^2)}\cos(\gamma_\mu Bt+\phi)$.
Here, $\gamma_\mu/2\pi= 135.5$~MHz/T is the muon gyromagnetic
ratio, $\phi$ is the initial phase of the muon-spin ensemble, and
$\sigma$ is the Gaussian relaxation rate. The right panel of
Fig.~\ref{fig:muSR}(a) shows that the TF asymmetry $A_0^{\rm TF}$
is almost temperature independent.  The slightly stronger
relaxation of the muon-spin polarization at 1.7~K relative to 20~K
is due to formation of the vortex lattice at $T<T_c$.

For the $x=0.45$ sample, Fig.~\ref{fig:muSR}(b), there is little
change in either the ZF or the TF time spectra on cooling from 20~K
to $\sim$7~K. At lower temperatures, however, an additional fast
relaxing component starts to develop. The solid lines in
Fig.~\ref{fig:muSR}(b) (left panel) correspond to fits with $A^{\rm
ZF}(t)=A_1^{\rm ZF} e^{-\Lambda_1^{\rm ZF} t}+A_2^{\rm ZF}
e^{-\Lambda_2^{\rm ZF} t}$ and $A^{\rm
TF}(t)=e^{-\sigma^2t^2/2}[A_1^{\rm TF}e^{-\Lambda_1^{\rm TF}
t}\cos(\gamma_\mu B_1t+\phi)+A_2^{\rm TF}e^{-\Lambda_2^{\rm TF}
t}\cos(\gamma_\mu B_2t+\phi)]$. Here, $A_{1(2)}^{\rm ZF(TF)}$ and
$\Lambda_{1(2)}^{\rm ZF(TF)}$ are the initial ZF(TF) asymmetry and
the exponentional depolarization rate of the slow (fast) relaxing
component, respectively. The decrease of $A_1^{\rm TF}$ with
decreasing temperature (Fig.~\ref{fig:muSR}(b) right panel) is due
to the development of magnetic order, which at $T\simeq1.7$~K
occupies more than 50\% of the whole sample volume. The LF
measurements reveal that the slow relaxing component completely
recovers at $\simeq10$~mT (similar to that observed for the $x=0.5$
sample), while the asymmetry of the fast relaxing one decreases by
$\sim50$\% at $B^{\rm LF}=0.4$~T. Considering that the muon spins
become decoupled from the static internal field $B_{\rm int}$ at
$B^{\rm LF}\gtrsim10B_{\rm int}$ \cite{Schenck86} we may assume that
the magnetism which develops in the $x=0.45$ sample below $T\sim7$~K
is caused by the static internal field $B_{\rm int}\gtrsim 0.1$~T at
the muon stopping site.

Magnetism was found to develop in the $x=0.4$, 0.25, 0.1, and 0.0
samples below $T\simeq18$, 30, 40 and 70~K, respectively, as
signalled by a fast drop of both $A^{\rm ZF}$ and $A^{\rm TF}$
within the first 100~ns [see Figs.~\ref{fig:muSR}(c) and (d); the ZF
and TF time spectra for $x=0.4$ and $x=0.1$ look very similar to
that of the $x=0.25$ sample and are not shown]. The TF and ZF data for
$x=$ 0.4, 0.25 and 0.1 were fitted similarly to $x=0.45$. In order
to fit the highly damped oscillations observed for $x=0$
[Fig.~\ref{fig:muSR}(d)] the second term in $A^{\rm ZF}(t)$ was
multiplied by  $\cos(\gamma_\mu B_{\rm int}t)$. This sample shows an
abrupt change in $B_{\rm int}\simeq 0.21$~T at $T\simeq70.6$~K and
hysteresis in $A_1^{\rm TF}(T)$ measured on increasing and
decreasing temperature --- see inset in the right panel of
Fig.~\ref{fig:muSR}(d). These features are evidence for a
first-order magnetic transition in Fe$_{1.03}$Te, consistent with
the results of Refs.~\onlinecite{Li09} and \onlinecite{Bao09}. To
within our experimental accuracy there is no hysteresis in the
magnetic transition for the $x=0.45$, 0.4, 0.25, and 0.1 samples.

\begin{figure}
\includegraphics[width=0.8\linewidth]{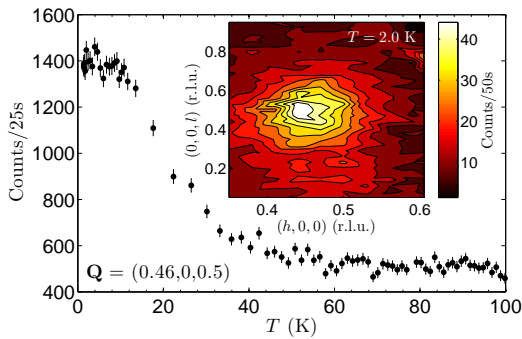}
 \vspace{-0.3cm}
\caption{\label{fig33} (Color online) Neutron diffraction from
Fe$_{1.03}$Se$_{0.25}$Te$_{0.75}$. The contour plot maps the
spin-flip scattering of polarized neutrons in the $(h,0,l)$ plane,
and reveals an incommensurate magnetic peak at $(0.46, 0, 0.5)$. The
main plot shows the temperature dependence of the intensity at
$(0.46, 0, 0.5)$ measured with unpolarized neutrons. }
\end{figure}

The fact that the ZF time spectra for the $x=0$ sample can be well
described by a damped cosine function with {\it zero} initial
phase [see Fig.~\ref{fig:muSR}(d)] suggests that the magnetism in
Fe$_{1.03}$Te is commensurate.\cite{Savici02} The absence of ZF
oscillations for samples with $x>0$ prevents any firm conclusions
being drawn about the type of magnetism in these samples. To learn
how the magnetic correlations change with Se doping we performed
neutron diffraction measurements on the $x=0.25$ crystal. The
neutron polarization analysis device MuPAD~\cite{mupad} was
employed to separate magnetic from non-magnetic scattering. The
inset in Fig.~\ref{fig33} is a color map of the spin-flip (SF)
scattering in the $(h, 0, l)$ plane in reciprocal space (referred
to the tetragonal unit cell with $a \simeq3.8$~{\AA} and $c
\simeq6.2$~{\AA} \cite{Bao09}). The sample temperature was 2~K,
and the neutron wavelength was 3.2~{\AA}. The neutron polarization
$\bf P$ was maintained parallel to the scattering vector $\bf Q$
so that the SF scattering is purely magnetic. The map reveals a
magnetic peak centered on the incommensurate wavevector
$(0.46,0,0.5)$. The peak is broader than the instrumental
resolution, and correlation lengths extracted from cuts through
the peak parallel to the $a$ and $c$ axes are $\xi_a = 11.5 \pm
1.0$ {\AA} and $\xi_c = 6.0 \pm 0.5$ {\AA}. Figure \ref{fig33}
shows the temperature dependence of the peak intensity at $(0.46,
0, 0.5)$ measured with unpolarized neutrons. The magnetic peak
emerges below $T \approx 40$ K, consistent with the muon asymmetry
data, Fig.~\ref{fig:muSR}(c).
A recent study on a crystal of
Fe$_{1.07}$Se$_{0.25}$Te$_{0.75}$ has also reported incommensurate
magnetic order.\cite{Wen09} The incommensurability and $\xi_c$ are consistent
with our results, but $\xi_a$ is a factor of 2 smaller than in our
sample.

\begin{figure}[htb]
\includegraphics[width=0.95\linewidth]{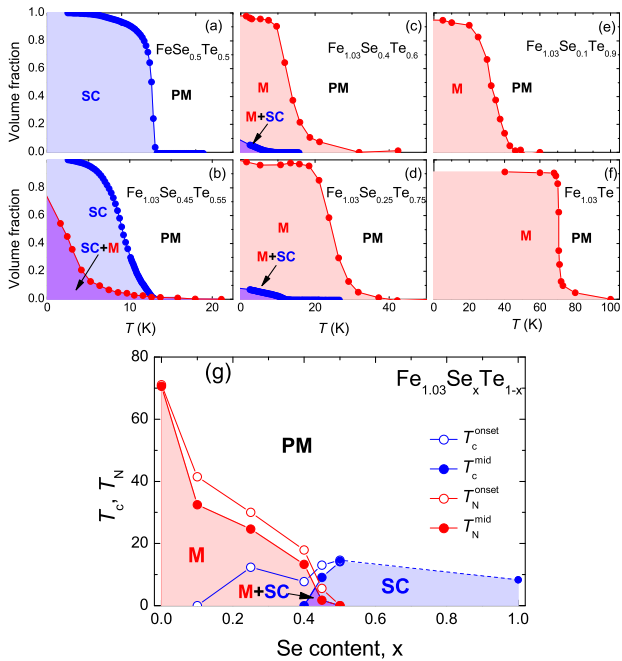}
 \vspace{-0.3cm}
\caption{(Color online) (a)--(f) Temperature dependence of the
superconducting (SC) and magnetic (M) volume fractions in
Fe$_{1.03}$Se$_{x}$Te$_{1-x}$. PM denotes the paramagnetic phase.
(g) Phase diagram showing $T_{\rm c}^{\rm onset}$, $T_{\rm c}^{\rm
mid}$ $T_{\rm N}^{\rm onset}$, and $T_{\rm N}^{\rm mid}$ as a
function of $x$. The datum for FeSe$_{1-x}$ is from
Ref.~\onlinecite{Khasanov08_FeSe}.}
 \label{fig:phase-separation}
\end{figure}

Figure~\ref{fig:phase-separation} summarizes our results on the
magnetism and superconductivity in Fe$_{1+y}$Se$_x$Te$_{1-x}$. The
volume fraction curves for the superconducting (SC) and magnetic (M)
phases are taken from $\chi_{\rm AC}(T)$
(Fig.~\ref{fig:magnetization}) and $A_1^{\rm TF}(T)$
(Fig.~\ref{fig:muSR}), respectively. The latter represents the
fraction of muons experiencing a static local field.
Figure~\ref{fig:phase-separation}(g) shows the mid-point and onset
of the superconducting and magnetic transitions, determined as shown
in Figs.~\ref{fig:magnetization} and \ref{fig:muSR}, as a function
of Se content $x$. It can be seen that superconductivity occurs
throughout the bulk of the $x=0.45$ and 0.5 crystals, while it
occupies up to $\simeq10$\% of the sample volume in the $x=0.25$ and
$x=0.4$ samples as $T\rightarrow 0$. Magnetic order is present in
the $x=0.45$, 0.4, 0.25, 0.1, and 0.0 samples with respective volume
fractions $\simeq75$\%, $98$\%, 98\%, 95\%, and 92\% at $T\rightarrow
0$.

Most interestingly, superconductivity and magnetism are shown to
coexist within certain temperature ranges in the $x=0.45$, 0.4 and
0.25 samples. For $x=0.45$, magnetism starts to develop below the
superconducting transition temperature
[Fig.~\ref{fig:phase-separation}(b)], while in $x=0.4$ and 0.25
magnetism appears first and superconductivity emerges at a lower
temperature [Figs.~\ref{fig:phase-separation}(c) and (d)]. The data
do not show any evidence that one form of order emerges at the
expense of the other, for if that were the case then a growth in one
order parameter would coincide with a decrease in the other. This is
clearly not the case, as may be seen in
Figs.~\ref{fig:phase-separation}(b)--(d). Nor do the data provide
any evidence for macroscopic phase separation into superconducting
and magnetic clusters (bigger than a few nm in size), as observed
e.g.\ for Ba$_{1-x}$K$_x$Fe$_2$As$_2$.\cite{Park09} In such a case
the sum of magnetic and superconducting volume fractions at a given
$T$ should never exceed unity as they do here, especially for
$x=0.45$.

To account for the coexistence of superconductivity and magnetism in
Fe$_{1+y}$Se$_x$Te$_{1-x}$ we consider two scenarios. The first
possibility is a {\it nanoscale} segregation into magnetic domains,
similar to that reported for cuprate HTS.\cite{Sanna04,
Savici02,Russo07} In underdoped cuprate HTS, static, short-range,
stripe-like magnetic correlations are thought to exist in the
superconducting state and are assumed not to affect the
superconducting carriers.\cite{Sanna04} Muons are sensitive to
dipolar fields at a distance of up to a few lattice spacings, so if
nano-scale magnetic domains exist then the fraction of muons
experiencing static local magnetic fields could be significantly
higher than the fraction of Fe sites carrying an ordered moment. On
the other hand, no evidence has been found yet for local magnetic
domains in Fe-based compounds. In fact, a recent nuclear magnetic
resonance study  of Ba(Fe$_{1-x}$Co$_x$)$_2$As$_2$
showed the appearance of magnetic order on {\it all} Fe sites thus
ruling out nano-scale segregation in that material.\cite{Laplace09}

The second possibility is a  coexistence of the two order
parameters on the {\it atomic} scale. The combination of incommensurate magnetism and
superconductivity is compatible with models recently proposed in Refs.~\onlinecite{Vorontsov09,Cvetkovic09}. According to Ref.~\onlinecite{Vorontsov09},
when $T_{\rm N}^{0}/T_{\rm c}^{\rm max}\sim 1$, where $T_{\rm N}^0$ is the
magnetic ordering temperature at zero doping
($x=0$) and $T_{\rm c}^{\rm max}$ is the maximum value of the of the superconducting transition temperature for a given family of Fe-based HTS,   the magnetic order is commensurate and the transition
between the magnetic and superconducting phases with $x$ is first
order. However, for larger $T_{\rm N}^{0}/T_{\rm c}^{\rm max}$ the transition
between commensurate magnetic order and superconductivity goes
through a region of $x$ where superconductivity coexists with
incommensurate magnetic order.  In the series
Fe$_{1+y}$Se$_x$Te$_{1-x}$ studied here we find just such behavior.
It is a commensurate magnet without superconductivity at $x=0$, and
a non-magnetic superconductor at $x=0.5$. In between, at $x=0.25$,
we observe incommensurate magnetism coexistent with $\sim$ 10\%
superconducting fraction. These results are encouraging for the
model, but details still need to be worked out. For example, both
LaFeAsO$_x$F$_{1-x}$ and Fe$_{1+y}$Se$_x$Te$_{1-x}$ have
 $T_{\rm N}^{0}/T_{\rm c}^{\rm max}\simeq 5$, yet LaFeAsO$_x$F$_{1-x}$ apparently exhibits a first-order
transition between magnetic and superconducting phases as a function
of $x$ without an intermediate region of coexistence.\cite{Luetkens09}

In conclusion, the phase diagram of Fe$_{1+y}$Se$_{x}$Te$_{1-x}$
bears a strong resemblance to that of other iron pnictide
superconductors, but the existence of an intermediate range of
doping in which superconductivity coexists with incommensurate
magnetic order appears to be specific to
Fe$_{1+y}$Se$_{x}$Te$_{1-x}$. The existence of such a phase has been
predicted theoretically and is of particular interest in view of the
possibility of a Fulde-Ferrell-Larkin-Ovchinnikov (FFLO) state.\cite{Cvetkovic09}

This work was performed at the S$\mu$S and SINQ,  Paul
Scherrer Institute (PSI, Switzerland). The work of MB was supported by the Swiss National Science Foundation. The work of EP was
supported by the NCCR program MaNEP. ATB thanks the PSI for support during an extended visit in 2009. RK acknowledges discussion with A.B.~Vorontsov.

\end{document}